\documentclass[a4paper,11pt]{article}

\usepackage{jheppub}
\usepackage{amsmath,amssymb,amsthm}
\usepackage{graphicx}
\usepackage{booktabs}
\usepackage{orcidlink}
\usepackage{comment}
\usepackage{xcolor}
\definecolor{red}{named}{black}

\long\def\colour#1#2{#2}
\long\def\new#1{{\color{red}#1}}
\newtheorem{theorem}{Theorem}

\newcommand{\HH}{\mathbb{H}}
\newcommand{\DD}{\mathbb{D}}
\newcommand{\RR}{\mathbb{R}}
\newcommand{\im}{\mathrm{Im}\,}
\newcommand{\re}{\mathrm{Re}\,}
\newcommand{\dd}{\mathrm{d}}

\title{A Bound on  the Dynamical Love Number}

\author[b]{Alex Kehagias\,
\orcidlink{0000-0001-6080-6215}
}
\author[a]{and Antonio Riotto\,
\orcidlink{0000-0001-6948-0856}
}

\affiliation[a]{D\'epartement de Physique Th\'eorique and Gravitational Wave Science Center,\\
Universit\'e de Gen\`eve, 24 quai E.\ Ansermet, CH-1211 Gen\`eve 4, Switzerland}
\affiliation[b]{Physics Division, National Technical University of Athens,\\
15780 Zografou Campus, Athens, Greece}

\emailAdd{kehagias@central.ntua.gr}
\emailAdd{antonio.riotto@unige.ch}


\abstract{The tidal deformability of a compact object is encoded in a single function of
frequency, the retarded Green's function relating the induced multipole to the applied
tide. Causality, reality, passivity, and the high-frequency conditions required for a
positive-measure dispersion representation allow this response to be rescaled into a
holomorphic self-map of the upper complex frequency half-plane. Using the 
Schwarz--Pick theorem, already employed to derive the  quantum chaos bound in black hole physics, we provide a bound  on the rate of the tidal response with respect to the frequency. For a neutron star with a positive tidal spectral threshold, the
dynamical Love number is bounded in terms of the static response and
the lowest frequency carrying nonzero tidal spectral weight, with the
ideal single-mode model saturating the bound. \new{For a black hole, whose tidal spectrum is gapless, the dynamical Love number is a finite-part moment of the horizon absorption spectrum. Under a growth hypothesis on the complete response the theorem constrains the full Euclidean response, but it does not bound the dynamical Love number.}}

\begin{document}
\maketitle
\flushbottom

\section{Introduction and summary of the results}
\label{sec:intro}

When a compact object, a neutron star or a black hole, sits in an external gravitational
field that varies across its size, it deforms and acquires induced multipole moments. The
proportionality between the applied tide and the induced moment defines the tidal Love
numbers, which imprint themselves on the gravitational waveform of a coalescing
binary~\cite{Flanagan:2007ix,Hinderer:2007mb} and are by now among the cleanest probes of
the internal structure of the inspiralling
bodies~\cite{Flanagan:2007ix,Rodriguez:2026iot,Chakraborty:2026qru}. A static tide
induces a constant moment; a time-dependent tide, as in any realistic inspiral, induces a
frequency-dependent one, whose low-frequency expansion defines the dynamical Love numbers.

Are these coefficients arbitrary, or does some principle constrain them?  The object that
carries the physics is the retarded Green's function $\chi(\omega)$ relating the induced
multipole to the applied tide. Viewed as a function of complex frequency it is far from
arbitrary. In fact, causality makes it holomorphic in the upper half-plane, reality makes it real
on the imaginary axis, where it defines the Euclidean Love number
$\lambda_{\rm E}(\nu)\equiv\chi(i\nu)$,  and passivity, the statement that the body absorbs
energy from the tide rather than feeding energy into it, fixes the sign of the dissipative
part on the real axis. These three inputs, supplemented by the high-frequency subtractions
needed for a dispersion relation with a positive spectral measure, identify a suitably
rescaled $\chi$ as a Herglotz function~\cite{donoghue1974}, that is a holomorphic
map of the upper half-plane into itself. 

We argue that, provided the conditions above are satisfied,  the dynamical Love number is not arbitrary, but is limited by the classical Schwarz--Pick
theorem~\cite{ahlfors1973conformal}, according to which,  a holomorphic self-map of the half-plane can only
shrink its hyperbolic distance, never enlarge it. Along the imaginary axis that distance
is nothing but distance in $\ln\nu$, so the contraction collapses to a single master
inequality,
\begin{equation}
\left|\frac{\dd\ln\Phi}{\dd\ln\nu}\right|\le1,
\label{eq:master1}
\end{equation}
where $\Phi(\nu)$ is the rescaled Euclidean response,
\begin{equation}
\Phi=\nu\,\lambda_{\rm E}\quad(\lambda_0\neq0),
\qquad\qquad
\Phi=-\lambda_{\rm E}/\nu\quad(\lambda_0=0),
\label{eq:PhiPsiintro}
\end{equation}
for a body with, respectively, a finite, like a neutron star, or a vanishing static Love number $\lambda_0$, like a black hole. 

This line of reasoning has appeared before in physics, and it is worth recalling where. The
same Schwarz--Pick rigidity is the engine behind the quantum chaos bound of Maldacena,
Shenker and Stanford~\cite{Maldacena:2015waa}, saturated on black hole photon
rings~\cite{Giataganas:2026ctn}. There one studies an out-of-time-order correlator that,
at temperature $T$, is analytic and bounded in a strip of imaginary time; mapping the
strip onto the disk and applying the contraction shows that the correlator cannot grow too
quickly, which is precisely the Lyapunov bound $\lambda_L\le2\pi T$ (in units
$\hbar=k_B=1$). The chaos bound is thus, at bottom, a Schwarz--Pick bound on a thermal
correlator. We apply the same geometry to a different thermal correlator, the tidal
response read on the imaginary-frequency Matsubara axis, so that the chaos bound and the
present bound become two instances of one mathematical fact. The same rigidity underlies
the sum rules and broadband bounds for passive electromagnetic and acoustic
media~\cite{Gustafsson:2010,Bernland:2011,Cassier:2017}.

What separates a star from a black hole is, remarkably, only which coefficient leads, and
hence which rescaling of $\chi$ is the Herglotz representative. A neutron star typically
carries a non-vanishing static Love number, the representative is $\omega\chi$, and the
theorem constrains the conservative dynamical Love number $\lambda_2$. \new{For a four dimensional Schwarzschild black hole, the static Love
number vanishes and the leading response is dissipative. Under the
stated analyticity and sublinear growth assumptions, $\chi/\omega$
is Herglotz and $\lambda_2$ is a finite-part moment of the absorption
spectrum. Since $g(0)=\eta>0$, the inverse moment entering the stellar
bound diverges. Schwarz-Pick constrains the Euclidean response but
does not yield the stellar bound on $\lambda_2$. These conclusions
remain conditional because the growth assumption has not been
established for the black hole response.} Passivity is the one dynamical input common to both. For
stars it is non-negative entropy production, for horizons the area theorem, two faces of a
single statement about the sign of absorption.

\begin{table}[t]
\centering
\footnotesize
\begin{tabular}{lll}
\toprule
Symbol & Meaning & Defined in\\
\midrule
$\chi(\omega)$ & retarded tidal response, $Q_{ab}(\omega)=-\chi(\omega)E_{ab}(\omega)$ & Eq.~\eqref{eq:chidef}\\
$\lambda_0,\ \eta,\ \lambda_2,\ \eta_3$ & static Love number, tidal heating, dynamical Love & Eq.~\eqref{eq:chiexp}\\
& number, cubic dissipative coefficient & \\
$\lambda_{\rm E}(\nu)=\chi(i\nu)$ & Euclidean Love number, real for $\nu>0$ & Eq.~\eqref{eq:euclidean}\\
$\lambda_\infty=\re\chi(\infty)$ & instantaneous (contact) response & Eq.~\eqref{eq:moments}\\
$\dd\mu(\omega')$ & tidal spectral measure & Eq.~\eqref{eq:measure}\\
$F(\omega)=\omega\,\chi(\omega)$ & Herglotz representative for $\lambda_0\neq0$ (stars) & Sec.~\ref{sec:bound}\\
$G(\omega)=\chi(\omega)/\omega$ & Herglotz representative for $\lambda_0=0$ (black holes) & Sec.~\ref{sec:bound}\\
$\Phi(\nu)=\nu\lambda_{\rm E}(\nu)$, $\Psi(\nu)=-\lambda_{\rm E}(\nu)/\nu$ & Euclidean representatives & Eq.~\eqref{eq:PhiPsi}\\
$\omega_{\rm gap}$ & lower endpoint of the tidal spectral support & Eq.~\eqref{eq:edgebound}\\
$\omega_*=\sqrt{\lambda_0/\lambda_2}$ & effective spectral frequency & Eq.~\eqref{eq:omegastar}\\
\new{$g(\omega)=\im\chi(\omega)/\omega$} & \new{absorptive weight, $g(0)=\eta$} & \new{Sec.~\ref{sec:bh}}\\
\new{$\lambda_2(\nu)$} & \new{Euclidean dynamical Love coefficient (gapless case)} & \new{Eq.~\eqref{eq:lambda2nu}}\\
\bottomrule
\end{tabular}

\vspace{3mm}

\begin{tabular}{lll}
\toprule
Physical assumption & Mathematical consequence & Where\\
\midrule
causality: $G_R(\tau)=0$ for $\tau<0$ & $\chi$ holomorphic in $\HH$ & Sec.~\ref{sec:response}\\
reality: $G_R(\tau)\in\RR$ & $\chi(-\bar\omega)=\overline{\chi(\omega)}$;\ $\lambda_{\rm E}(\nu)\in\RR$;\ even/odd & Eq.~\eqref{eq:euclidean}\\
& split of conservative/dissipative terms & \\
passivity: second law, area theorem & $\omega\,\im\chi(\omega)\ge0$;\ measure $\dd\mu\ge0$ & Eqs.~\eqref{eq:passivity}, \eqref{eq:measure}\\
high-frequency control (one subtraction) & Kramers--Kronig; moment representation & Eqs.~\eqref{eq:kk}--\eqref{eq:euclidrep}\\
{\colour{blue}{conservative leading case with $\lambda_\infty\ge0$}} & representative Herglotz: $\im F\ge0$ on $\HH$ & Eqs.~\eqref{eq:lambdainftypos}--\eqref{eq:Fpositive}\\
all of the above & Schwarz--Pick: $\left|\dd\ln\Phi/\dd\ln\nu\right|\le1$ & Eqs.~\eqref{eq:twopoint}--\eqref{eq:master}\\
\bottomrule
\end{tabular}
\caption{Summary of the notation (top) and of the logical chain from physical assumptions
to the properties underlying the bound (bottom). All statements hold per multipole,
parity and, for rotating bodies, azimuthal channel.}
\label{tab:summary}
\end{table}

The plan is as follows. Section~\ref{sec:response} introduces the response
function and its three structural properties; Section~\ref{sec:bound} derives the master
inequality; Sections~\ref{sec:stars} and~\ref{sec:bh} treat neutron stars and black holes
in turn; Section~\ref{sec:conclusions} concludes; two Appendices  collect
the technical material. Finally, table~\ref{tab:summary} collects the notation used throughout
the paper and the chain leading from the physical assumptions to the mathematical
properties on which the bound rests.

\section{The tidal response function}
\label{sec:response}

In the worldline effective field theory the body is a point particle carrying finite-size
operators built from the curvature on its
worldline~\cite{Goldberger:2004jt,Hinderer:2007mb,Kol:2011vg}. For the dominant
electric-type quadrupole the external tide is the electric part of the Weyl tensor
projected on the body's frame, $E_{ab}=C_{0a0b}|_{\rm worldline}$. 
The full action $S$ contains an expansion in $E_{ab}$ and its proper-time derivatives, a
linear coupling of the tide to the internal structure, and the action $S_{\rm internal}$
of the internal degrees of freedom themselves,
\begin{equation}
S\ \supset\ \int \dd\tau\left[\tfrac14\,\mu^{(0)}_{\rm E}\,E_{ab}E^{ab}
+\tfrac14\,\mu^{(2)}_{\rm E}\,\dot E_{ab}\dot E^{ab}+\dots
-\tfrac12\,\hat Q^{ab}_{\rm int}\,E_{ab}\right]+S_{\rm internal}.
\label{eq:Sfs}
\end{equation}
Here $\hat Q^{ab}_{\rm int}$ is the multipole operator, a composite built from the
genuinely dynamical internal degrees of freedom of the body, i.e., the fluid oscillation
modes for a star and the horizon degrees of freedom for a black hole, whose free dynamics
is collected in $S_{\rm internal}$. The
local terms, with Wilson coefficients $\mu^{(n)}_{\rm E}$, encode the conservative part
of the response, while the coupling to $\hat Q^{ab}_{\rm int}$ supplies absorption. The
total induced quadrupole is the quantity conjugate to the applied tide,
$Q_{ab}=-\delta S/\delta E_{ab}$, in the same way that a magnetization is conjugate to an
applied magnetic field. Switching on a weak tide and evaluating $\langle
Q_{ab}\rangle$ in first-order (in-in) perturbation theory gives the Kubo formula of
linear response, 
\begin{equation}
    Q_{ab}(t)=-\int_{-\infty}^{t}\dd t'\,G_R(t-t')\,E_{ab}(t'),
\end{equation}
where $G_R(\tau)$ is 
the retarded Green's function
\begin{equation}
G_R(\tau)=i\theta(\tau)\langle[\hat Q_{ab}(\tau),\hat Q^{ab}(0)]\rangle.
\end{equation}
In frequency space the response function reads
\begin{equation}
\chi(\omega)=\int_0^\infty \dd\tau\,e^{i\omega\tau}\,G_R(\tau),
\qquad Q_{ab}(\omega)=-\chi(\omega)\,E_{ab}(\omega),
\label{eq:chidef}
\end{equation}
which is the gravitational analogue of the polarizability of an atom. The local terms
of~\eqref{eq:Sfs} contribute its analytic, conservative pieces, while the internal
dynamics supplies its dissipative ones. The low-frequency expansion
\begin{equation}
\chi(\omega)=\lambda_0+i\,\eta\,\omega+\lambda_2\,\omega^2+i\,\eta_3\,\omega^3+\dots
\label{eq:chiexp}
\end{equation}
defines the static Love number $\lambda_0$, the tidal-heating coefficient $\eta$, the
conservative dynamical Love number $\lambda_2$, and so
on~\cite{Goldberger:2005cd,Damour:2009vw,Binnington:2009bb}. The identical construction
runs independently at every multipole $\ell$, parity and, for a rotating body, azimuthal
number $m$. All three properties below act within a single channel, so every bound in
this paper holds per multipole, and we suppress the label except where it matters.

The analytic structure of the response follows from three properties of the retarded Green's function. First, causality: $G_R$ vanishes at negative times, so its Fourier transform $\chi$ is holomorphic in the upper complex half-plane
$$\HH=\{\omega\in\mathbb{C}\mid\im\omega>0\}.$$ (see, e.g.,  Ref. \cite{Correia:2025enx}).
Second, reality: $G_R$ is real in the time domain, so its transform obeys the reflection symmetry $\chi(-\bar\omega)=\overline{\chi(\omega)}$. In particular $\chi$ is real along the positive imaginary axis, which permits the definition of the Euclidean Love number
\begin{equation}
  \lambda_{\rm E}(\nu)\equiv\chi(i\nu)\in\RR\qquad(\nu>0).  \label{eq:euclidean}
\end{equation}
Reality also forces the conservative coefficients in Eq.~\eqref{eq:chiexp} to multiply even powers of $\omega$ and the dissipative coefficients to multiply odd powers. Third, passivity restricts the energy exchange the response can mediate. For a monochromatic tide, the cycle-averaged power absorbed by the body is 
\begin{eqnarray}
    \langle\dot W\rangle=\tfrac12\,\omega\,\im\chi(\omega)\,|E|^2.
\end{eqnarray}
A passive system cannot deliver net energy to the external field, so $\langle\dot W\rangle\geq 0$ and
\begin{equation}
 \omega\,\im\chi(\omega)\ \ge\ 0\qquad(\omega\in\RR). 
\label{eq:passivity}
\end{equation}
Let us pause on the physical mechanism behind this condition, since its immediate
manifestation differs between stars and black holes even though it stems from one law. For
a horizonless stellar body the oscillating tide excites internal degrees of freedom, fluid
modes and shear stresses, which couple to viscosity and radiative losses. The energy fed
in is irreversibly converted to heat at a non-negative rate, so entropy production is
non-negative by the second law. The absorptive part of the response is exactly this
dissipation rate per unit driving, and the fluctuation--dissipation theorem identifies
$\im\chi(\omega)$ with the spectral density of the body's equilibrium multipole
fluctuations, which is non-negative. A body with $\im\chi<0$ over some band would amplify
the tide rather than damp it, the hallmark of an active medium and impossible in stable
thermal equilibrium~\cite{Nussenzveig}. For a black hole the same inequality has a
gravitational origin. Because of the presence of the horizon  energy that crosses it does not
return, which is what makes the body a perfect passive absorber. The flux the tide drives
across the horizon is governed by the first law,
\begin{equation}
\dd M=\frac{\kappa}{8\pi}\,\dd A+\Omega_H\,\dd J,
\end{equation}
and Hawking's area theorem, $\dot A\ge0$ for matter obeying the null energy condition,
gives $\im\chi(\omega)$ the sign of $\dot A$, so that~\eqref{eq:passivity} becomes the
second law of black hole mechanics.

One subtlety arises for rotating black holes. The frequency
in~\eqref{eq:passivity} is the one measured in the horizon-corotating frame,
$\tilde\omega=\omega-m\Omega_H$ for azimuthal number $m$, and passivity holds in terms of
$\tilde\omega$. For $0<\omega<m\Omega_H$ the horizon gives up rotational energy to the
wave, superradiance, and the lab-frame $\im\chi$ turns negative, yet the area theorem
survives because the corotating flux keeps its sign.

For a non-rotating body, or after decomposing the real tidal field into a real harmonic
basis, the independent response channels contribute additively to the cycle-averaged work,
\begin{equation}
\langle\dot W\rangle
=\frac12\sum_{\ell,A}\omega\,\im\chi_{\ell A}(\omega)
\,|E_{\ell A}|^2,
\label{eq:workmultipole}
\end{equation}
where $A$ labels the independent real harmonic components and parities. Positivity for an
arbitrary excitation then requires the corresponding absorptive quadratic form to be
non-negative in every channel. 

Turning this boundary passivity into an interior Herglotz property also requires control
of the response at large complex frequency (see also Ref. \cite{Creminelli:2025rxj}). {\colour{blue}{Causality and passivity do not by themselves imply that $\chi(z)$ approaches a finite constant at complex infinity, and the point particle EFT does not determine the arbitrarily high frequency response after its long wavelength expansion has ceased to apply. The existence of $\lambda_\infty$ and the one subtraction form used below are therefore additional hypotheses about the complete matched susceptibility rather than consequences of the low energy theory. This issue is familiar in gravitational dispersion relations and is closely related to the subtraction ambiguities discussed in Refs.~\cite{Porto:2016pyg,Goldberger:2020fot}. When further subtractions are required the associated polynomial terms must be retained, while if the Herglotz property is established only below a finite complex frequency scale, the finite domain result of subsection~\ref{sec:finite-domain} is the appropriate statement. The black hole case is treated with this qualification throughout Section~\ref{sec:bh} and the formal large frequency obstruction of the scheme independent skeleton is discussed in subsection~\ref{sec:bhscope}. We do not presently have a first principles determination of
$\chi(\infty)$ from the ultraviolet behavior of the full gravitational two
point function and we do not assume that such a determination is available.
Our result is therefore conditional on the stated subtraction and high
frequency hypothesis.}} We assume that the complete, matched
susceptibility, as opposed to an isolated local Wilson coefficient in a fixed scheme,
admits the subtraction at infinity used below. In particular, $\chi(z)-\lambda_\infty$ decays
sufficiently fast in the upper half-plane, with no additional polynomial subtractions\footnote{Usually, if the  contribution
from the large semi-circle does not vanish at complex infinity in Green's functions, one can  adopt the so-called  modified dispersion relations  going to zero at complex infinity, possibly implemented by  divisions by a polynomial. We will come back to this point in \new{subsection~\ref{sec:subtracted}} when dealing with black holes.
} Under this assumption, a Cauchy contour in the upper half-plane yields the
Kramers--Kronig relation
\begin{equation}
\re\chi(\omega)-\re\chi(\infty)=\frac{1}{\pi}\,\mathrm{P}\!\!\int_{-\infty}^{\infty}\frac{\im\chi(\omega')}{\omega'-\omega}\,\dd\omega'. 
\label{eq:kk}
\end{equation}
Folding the integration domain onto the positive real axis with the reality condition and expanding the integrand in powers of $\omega^2$ maps each conservative coefficient onto a moment of the dissipation spectrum. Defining the non-negative measure
\begin{equation}
  \dd\mu(\omega')=\frac{2}{\pi}\,\frac{\im\chi(\omega')}{\omega'}\,\dd\omega'\ \ge\ 0,  \label{eq:measure}
\end{equation}
whose positivity is enforced by passivity, we obtain
\begin{equation}
\lambda_0=\lambda_\infty+\int_0^\infty\dd\mu(\omega'),\qquad\lambda_{2n}=\int_0^\infty\frac{\dd\mu(\omega')}{\omega'^{\,2n}}\quad(n\ge1)  \label{eq:moments}
\end{equation}
Here $\lambda_\infty=\re\chi(\infty)$ is the instantaneous component of the response and carries no spectral weight. The Love numbers are therefore not independent data, but successive moments of one positive measure, where a conservative quantity measured at low frequency knows about dissipation at every frequency. {\colour{blue}{In Eqs.~\eqref{eq:moments} and in the stellar bounds below the coefficients $\lambda_{2n}$ denote coefficients of the complete matched susceptibility after the subtraction polynomial has been fixed. An isolated renormalized worldline coefficient can shift under a finite counterterm and is not separately constrained by Schwarz-Pick. The eligible object is the complete response or an equivalent combination in which the subtraction scale cancels.}} On the Euclidean axis the same
dispersion relation reads
\begin{equation}
\lambda_{\rm E}(\nu)=\lambda_\infty+\int_0^\infty
\frac{\omega'^2}{\omega'^2+\nu^2}\,\dd\mu(\omega'),
\label{eq:euclidrep}
\end{equation}
manifestly real, monotonically decreasing in $\nu$, and interpolating between $\lambda_0$
at $\nu=0$ and $\lambda_\infty$ at $\nu\to\infty$. The Poisson kernel smears the boundary
dissipation into the interior analytic quantity on which the contraction theorem acts.
Notice that the sign of the instantaneous term matters, since only for
\begin{equation}
\lambda_\infty\ge0
\label{eq:lambdainftypos}
\end{equation}
does the representation \eqref{eq:euclidrep} directly establish that $F(z)=z\chi(z)$ is Herglotz. Indeed,
for $z=x+iy$ with $y>0$,
\begin{align}
F(z)&=\lambda_\infty z+
\int_0^\infty\frac{z\,\Omega^2}{\Omega^2-z^2}\,\dd\mu(\Omega),\\
\im F(z)&=\lambda_\infty y+
\int_0^\infty
\frac{\Omega^2 y\left(\Omega^2+|z|^2\right)}
{|\Omega^2-z^2|^2}\,\dd\mu(\Omega)\ge0.
\label{eq:Fpositive}
\end{align}
Thus the Herglotz property follows from the positive dispersion representation and
$\lambda_\infty\ge0$, rather than from boundary passivity alone. Strict positivity of
$\lambda_\infty$ is not required since $\lambda_\infty=0$ is admissible whenever the spectral
measure is non-zero. {\colour{blue}{The condition $\lambda_\infty\ge0$ is specific to the conservative leading representative $F=z\chi$ and must not be imposed in the black hole case. If the same one subtraction representation is valid and the static Love number vanishes, then Eq.~\eqref{eq:moments} instead gives
\begin{equation}
\lambda_\infty=-\int_0^\infty\dd\mu(\Omega)<0
\label{eq:bhlambdainfty}
\end{equation}
for nonzero absorption. Inserting this relation into the dispersive representation yields
\begin{equation}
\chi(z)=z^2\int_0^\infty\frac{\dd\mu(\Omega)}{\Omega^2-z^2},
\qquad
G(z)=\frac{\chi(z)}{z}
=z\int_0^\infty\frac{\dd\mu(\Omega)}{\Omega^2-z^2}.
\label{eq:Gfrommeasure}
\end{equation}
For $z=x+iy$ with $y>0$ one then obtains
\begin{equation}
\im G(z)
=
y\int_0^\infty
\frac{\Omega^2+|z|^2}{|\Omega^2-z^2|^2}\,\dd\mu(\Omega)
\ge0,
\label{eq:Gpositive}
\end{equation}
so $G$ is Herglotz whenever the assumed dispersion representation is valid. On the Euclidean axis the same identity gives
\begin{equation}
\lambda_{\rm E}(\nu)
=
-\nu^2\int_0^\infty\frac{\dd\mu(\Omega)}{\Omega^2+\nu^2}<0,
\qquad
\Psi(\nu)
=
\nu\int_0^\infty\frac{\dd\mu(\Omega)}{\Omega^2+\nu^2}>0.
\label{eq:BHsigns}
\end{equation}
Thus the negative subtraction constant required by a vanishing black hole static Love number is consistent with the Herglotz property of the different representative $G$, while the conclusion remains conditional on the validity of the dispersion representation itself.}} \new{The moment formula for $n\ge1$ in Eq.~\eqref{eq:moments}, and the expansion of Eq.~\eqref{eq:Gfrommeasure} in even powers of $z$, require interchanging the Taylor expansion with the spectral integral. This is legitimate for a measure with a positive threshold, but not for gapless dissipation. A non-vanishing term $i\eta\omega$ in Eq.~\eqref{eq:chiexp} requires the inverse moment $\int_0\dd\mu/\Omega^2$ to be infrared divergent, and it arises precisely from the non-commutativity of the expansion and the integral. This divergence is not a pathology but reflects gapless dissipation, and in that case the coefficients $\lambda_{2n}$ must be defined as finite-part moments, see subsection~\ref{sec:finitepart}.} If a physical problem requires further subtractions, the corresponding
polynomial terms must be included and their signs controlled before Schwarz--Pick can be
applied. \new{Polynomial terms of degree greater than one in the Herglotz
representative are incompatible with Eq.~\eqref{eqherglotz}.
For $G=\chi/z$, a term $bz^2$ in $\chi$ adds $bz$ to $G$ and
preserves its Herglotz property for $b\ge0$, but is excluded
by the sublinear growth assumption~\eqref{eq:sublinear}. The worldline EFT allows such contact terms and their finite parts are scheme dependent, so their absence in the complete response is a genuine assumption, not determined by the absorptive part. It holds by construction for the non-relativistic star, whose response is a convergent mode sum, but not in the relativistic case or for black holes, where our results are conditional on it.} 
The bound of the next section is, in one sentence, the statement that the
moments~\eqref{eq:moments} of a positive measure cannot be arbitrary and that high moments
are controlled by low ones.

\section{The Schwarz--Pick bound}
\label{sec:bound}

A Herglotz function is a function $F$ holomorphic on $\HH$ whose imaginary part is
non-negative there. Unless $F$ is a real constant, its imaginary part is a non-negative
harmonic function that cannot be zero in the interior by the minimum principle, so a
Herglotz function is a genuine holomorphic map of $\HH$ into itself. Every function of this class can be represented by a positive measure,
yielding a manifestly positive form of the dispersion relation. 
For a causal response with the growth and subtraction properties stated in
Section~\ref{sec:response}, the positive spectral representation translates physical
passivity into the Herglotz property~\cite{Akhiezer,Nussenzveig,Garnett}.
A related theorem, the
Schwarz--Pick theorem \cite{ahlfors1973conformal},
states that a holomorphic
map of $\HH$ into itself can only contract the hyperbolic metric $\dd s=|\dd w|/\im w$
and can never stretch it. Two geometric features of this metric do all the work. The boundary $\im w=0$ lies at infinite hyperbolic distance, so no non-constant self-map can push an interior point onto the real axis. The imaginary axis is a geodesic with length element $\dd\nu/\nu$, so hyperbolic distance along it is simply distance in $\ln\nu$. For a Herglotz function that preserves the imaginary axis, $F(i\nu)=i\,\Phi(\nu)$ with $\Phi>0$, the contraction therefore reads

\begin{equation}
\big|\ln\Phi(\nu_1)-\ln\Phi(\nu_2)\big|\ \le\ \big|\ln\nu_1-\ln\nu_2\big|\label{eq:twopoint}.
\end{equation}
The differential form of this relation yields the master inequality~\eqref{eq:master1}, i.e., 

\begin{equation}
\left|\frac{\dd\ln\Phi}{\dd\ln\nu}\right|\le1,  
\label{eq:master}
\end{equation}
the statement that the logarithmic slope of an admissible response along the Euclidean
axis can never exceed unity in magnitude. Equality at an interior point requires the map
to be a M\"obius automorphism of the upper half-plane, hence a global hyperbolic
isometry. Appropriately phased power laws $F(\omega)\sim \omega^s$ provide useful
examples with constant logarithmic slope in the allowed interval, but for $|s|<1$ they
are strict contractions, not automorphisms.

The response function $\chi(\omega)$ itself need not belong to the Herglotz class, because
its absorptive part changes sign under $\omega\to-\omega$. The infrared behavior suggests
a natural candidate rescaling, but the Herglotz property must then be verified for the
complete response using its dispersion representation and high-frequency behavior. 

For a
system with a non-vanishing static response, like neutron stars, the representative
$F(\omega)=\omega\chi(\omega)$ is Herglotz under the assumptions leading to
Eqs.~\eqref{eq:lambdainftypos}--\eqref{eq:Fpositive}. {\colour{blue}{When $\lambda_0=0$ and the leading response is dissipative, $G(\omega)=\chi(\omega)/\omega$ is regular at the origin. If the one subtraction positive measure representation of Section~\ref{sec:response} is valid, Eqs.~\eqref{eq:bhlambdainfty} and \eqref{eq:Gpositive} prove that $G$ is Herglotz even though $\lambda_\infty<0$. If that representation is not known globally, as in the present black hole problem, the Schwarz-Pick statement remains conditional and only the finite domain form can be used in a region where the Herglotz property has independently been established. Boundary positivity by itself is not sufficient.}} On the Euclidean
axis the two choices give
\begin{equation}
\Phi(\nu)\equiv\nu\,\lambda_{\rm E}(\nu)\quad(\lambda_0\neq0),
\qquad\qquad
\Psi(\nu)\equiv-\frac{\lambda_{\rm E}(\nu)}{\nu}\quad(\lambda_0=0),
\label{eq:PhiPsi}
\end{equation}
and the master inequality applies to each. A power-law response gives the representatives
a branch point at $\omega=0$, which sits on the boundary of $\HH$ and leaves the interior
single-valued and holomorphic. Notice also that the  inequality (\ref{eq:master})  constrains the complete, matched susceptibility, not the Taylor polynomial or an individual Wilson coefficient in a fixed renormalization scheme. 

\subsection{Finite-frequency domain}
\label{sec:finite-domain}
The bound~\eqref{eq:master} assumes that the relevant response is a
Herglotz function throughout the upper half-plane.  In some applications,
however, analyticity and positivity may be under control only inside a finite
complex-frequency domain.  Suppose, more precisely, that the representative
$\mathcal R(\omega)$ maps the upper half-disk
\begin{equation}
D_\Lambda
=
\left\{
\omega\in\mathbb C
\ \middle|\
\operatorname{Im}\omega>0,\quad |\omega|<\Lambda
\right\}
\end{equation}
holomorphically into the upper half-plane and satisfies
$\mathcal R(i\nu)=iY(\nu)$, with $Y(\nu)>0$.  Here $\Lambda$ denotes the
radius of the domain in the complex-frequency plane in which the Herglotz
property has been established, and it is not a renormalization scale.
The half-disk is mapped conformally onto the upper half-plane by
\begin{equation}
\mathcal C_\Lambda(\omega)
=
\left(
\frac{\Lambda+\omega}{\Lambda-\omega}
\right)^2.
\label{eq:halfdiskmap}
\end{equation}
Along the imaginary-frequency axis, the natural hyperbolic coordinate of
$D_\Lambda$ is
\begin{equation}
s_\Lambda(\nu)
=
\ln\left(
\frac{2\Lambda\nu}{\Lambda^2-\nu^2}
\right),
\qquad
0<\nu<\Lambda.
\label{eq:halfdiskcoordinate}
\end{equation}
The Schwarz--Pick contraction therefore gives
\begin{equation}
\left|
\frac{\mathrm d\ln Y}{\mathrm d\ln\nu}
\right|
\leq
\frac{\Lambda^2+\nu^2}{\Lambda^2-\nu^2},
\qquad
0<\nu<\Lambda.
\label{eq:finiteSP}
\end{equation}
The corresponding two-point inequality is
\begin{align}
\left|
\ln\frac{Y(\nu_2)}{Y(\nu_1)}
\right|
&\leq
\left|
s_\Lambda(\nu_2)-s_\Lambda(\nu_1)
\right|,
\end{align}
which is explicitly written as
\begin{align}
\left|
\ln\frac{Y(\nu_2)}{Y(\nu_1)}
\right|
&\leq
\left|
\ln\left[
\frac{\nu_2(\Lambda^2-\nu_1^2)}
     {\nu_1(\Lambda^2-\nu_2^2)}
\right]
\right|.
\label{eq:finiteSPtwopoint}
\end{align}
For frequencies well inside the controlled domain,
\begin{equation}
\frac{\Lambda^2+\nu^2}{\Lambda^2-\nu^2}
=
1+2\frac{\nu^2}{\Lambda^2}
+\mathcal O\left(\frac{\nu^4}{\Lambda^4}\right),
\end{equation}
and the global result~\eqref{eq:master} is recovered as
$\nu/\Lambda\to0$, or as $\Lambda\to\infty$ at fixed $\nu$.  The divergence
of Eq.~\eqref{eq:finiteSP} as $\nu\to\Lambda$ does not imply a singular
response.  It expresses the loss of constraining power as the boundary of the
domain in which the Herglotz property is known is approached.  Importantly,
Eq.~\eqref{eq:finiteSP} requires positivity throughout the complex half-disk, since
control of the response only along the imaginary-frequency axis is not
sufficient.
\section{Neutron stars}
\label{sec:stars}

In the stellar regime considered here the static Love number $\lambda_0$ is non-zero,
and, under the assumptions of Section~\ref{sec:response},
$\Phi=\nu\lambda_{\rm E}$ is the appropriate Herglotz representative. The Herglotz
property enforces $\Phi>0$, so the Euclidean Love number is strictly positive,
$\lambda_{\rm E}(\nu)>0$. Inserting the expansion
$\lambda_{\rm E}(\nu)=\lambda_0-\lambda_2\nu^2+\dots$, which follows from
Eqs.~\eqref{eq:chiexp} and \eqref{eq:euclidean}, the slope is
 
\begin{equation}
\frac{\dd\ln\Phi}{\dd\ln\nu}=1-\frac{2\lambda_2}{\lambda_0}\,\nu^2+\dots\, .
\label{eq:slopestar}
\end{equation}
At this order, the upper side of the master inequality implies
$\lambda_2\ge0$. The lower side gives, at a fixed sufficiently small $\nu$, only the
truncated consistency condition
\begin{equation}
\lambda_{2,\ell}\ge0,
\qquad
\lambda_{2,\ell}\lesssim\frac{\lambda_{0,\ell}}{\nu^2},
\label{eq:starbound}
\end{equation}
within the displayed low-frequency truncation.
The second relation diverges as $\nu\to0$ and is not a finite upper bound on the Taylor
coefficient. The robust local information is the sign $\lambda_{2,\ell}\ge0$. A finite
magnitude bound requires additional spectral information.

If we define the lower endpoint of the positive spectral measure in the chosen channel by $\omega_{\rm gap}$, then 
for $\omega_{\rm gap}>0$, the moment representation gives the primary finite bound
\begin{equation}
\lambda_{2,\ell}\le
\frac{\lambda_{0,\ell}-\lambda_{\infty,\ell}}
{\omega_{{\rm gap},\ell}^{2}}.
\label{eq:edgebound}
\end{equation}
Each parameter entering this inequality is, in principle, independently
testable. The static deformability can be extracted from the inspiral
tidal phasing, while the characteristic stellar-mode frequency may be
constrained through gravitational-wave asteroseismology or empirical
relations involving the stellar mass and radius
\cite{Andersson:1997rn}.
The dynamical correction is encoded in the frequency dependence of the
tidal response.

The inequality \eqref{eq:edgebound} follows directly from the spectral support, without
evaluating a truncated Taylor series at the boundary of its convergence disk 
\begin{equation}
\lambda_2=\int\frac{\dd\mu(\omega')}{\omega'^{2}}
\le\frac{1}{\omega_{\rm gap}^{2}}\int\dd\mu(\omega')
=\frac{\lambda_0-\lambda_\infty}{\omega_{\rm gap}^{2}}.
\label{eq:momentbound}
\end{equation}
This oscillator-strength sum rule shows that only the spectral part
$\lambda_0-\lambda_\infty$ drives the dynamical correction. The strict threshold is the
lowest frequency with non-zero tidal spectral weight, not necessarily the most prominent
mode. In a cold, non-rotating, barotropic model it is commonly the fundamental $f$-mode,
so that $\omega_{\rm gap}=\omega_f$. On the other hand, in a stratified or rotating star it may instead be a
coupled $g$-mode, inertial mode, interface mode, or the edge of a continuum
\cite{Lai:1993di,Steinhoff:2016rfi,Hinderer:2016eia}. Let us stress that a horizonless object is not guaranteed to have a positive threshold,
since dissipative or gapless matter channels can extend the spectral support all the way
to zero. In that case the finite bound~\eqref{eq:edgebound} becomes unavailable, although
the full-function Schwarz--Pick inequality may still apply. \new{The moment expansion~\eqref{eq:moments}, and the bounds of this section, therefore apply to the ideal gapped response. For a dissipative stellar model with $\im\chi\simeq\eta\,\omega$~\cite{HegadeKR:2024slr}, the inverse moment diverges, $\lambda_2$ is the finite-part moment~\eqref{eq:lambda2fp} of Section~\ref{sec:bh}, and neither the finite bound~\eqref{eq:edgebound} nor the sign $\lambda_2\ge0$ is guaranteed.}

The inclusion of a characteristic scale within the finite moment bound is dimensionally
mandatory, because $\lambda_2/(\lambda_0-\lambda_\infty)$ has dimensions of time squared.
When a positive threshold exists, the positive-measure representation identifies the sharp
scale as $\omega_{\rm gap}$ rather than the stellar radius or an arbitrary microscopic
cutoff. The threshold may be inferred from a mode calculation, a spectral computation, or
suitable phenomenological information. With only this spectral input the bound is
optimal, where  a single mode carrying all the spectral weight at the threshold saturates
Eq.~\eqref{eq:edgebound}.

Indeed, this bound is saturated by the simplest stellar model. The single-mode response function
\begin{equation}
\lambda(\omega)=\frac{\lambda_0\,\omega_f^2}{\omega_f^2-\omega^2}
\label{eq:fmode}
\end{equation}
introduced in Refs.~\cite{Steinhoff:2016rfi,Hinderer:2016eia} (and well confirmed by perturbative calculations \cite{Apostolidis:2026qsg}) belongs to the Herglotz
class after appropriate rescaling. Its susceptibility has poles at $\pm\omega_f$, while
the folded positive-frequency measure consists of a single point mass at $\omega_f$. {\colour{blue}{
For the ideal single mode response in Eq.~\eqref{eq:fmode} one has
$\lambda(\omega)\to0$ as $|\omega|\to\infty$ and therefore
$\lambda_\infty=0$ by construction.
This is a property of the ideal single mode model and is not an assumption
that $\lambda_\infty$ vanishes for a generic relativistic neutron star.
}} The model yields $\lambda_2=\lambda_0/\omega_f^2$ exactly and
therefore saturates the spectral moment bound~\eqref{eq:edgebound}. Its Euclidean
logarithmic slope, $(\omega_f^2-\nu^2)/(\omega_f^2+\nu^2)$, varies monotonically
between the two limiting values and approaches the Schwarz--Pick boundaries only
asymptotically. 

\begin{figure}[t]
\centering
\includegraphics[width=\textwidth]{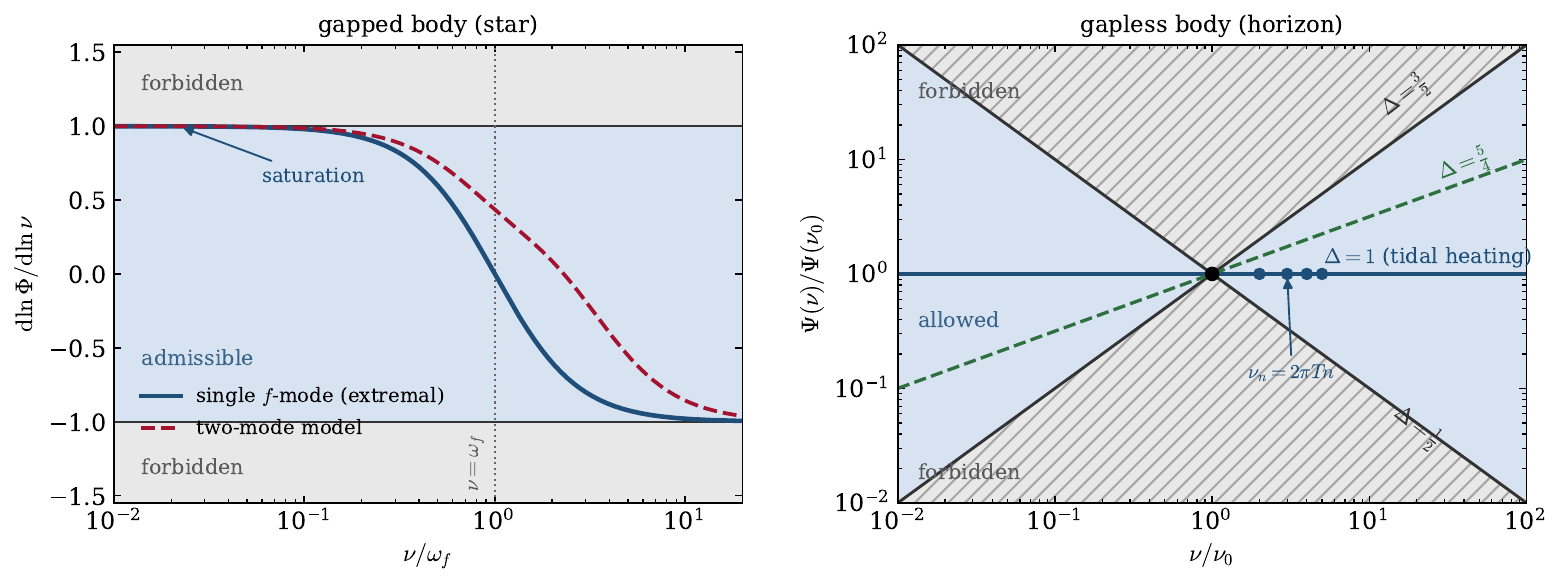}
\caption{
Constraints on compact object tidal responses derived from the Schwarz--Pick bound. The
left panel displays the logarithmic slope of the stellar representative
$\Phi=\nu\lambda_{\rm E}$ as a function of Euclidean frequency. The shaded band is the
admissible region $|\dd\ln\Phi/\dd\ln\nu|\le1$. The solid curve is the ideal single $f$-mode model, which saturates the spectral moment
bound and approaches the two local Schwarz--Pick boundaries only asymptotically, wheras the
dashed curve is a two-mode example. The vertical line $\nu=\omega_f$ marks the modal
scale of the ideal model, not a general identity between the spectral threshold and the
convergence radius. The right panel shows the geometric two-point wedge for an exact
black hole representative $\Psi=-\lambda_{\rm E}/\nu$, normalized at a reference scale
$\nu_0$. Power laws $\Psi\propto\nu^{2\Delta-2}$ trace straight lines inside the wedge.
The plotted Matsubara points are reference frequencies, and  they are not, by themselves,
boundaries of the domain of validity of a black hole approximation.}
\label{fig:bound}
\end{figure}

Beyond low-frequency expansions, the integral representation~\eqref{eq:euclidrep} yields exact constraints valid at every frequency. It shows at once that the Euclidean Love number is squeezed between its asymptotic limits,
\begin{equation}
\lambda_\infty\ \le\ \lambda_{\rm E}(\nu)\ \le\ \lambda_0
\qquad\text{for all }\nu>0.
\label{eq:sandwich}
\end{equation}
Globally, the master inequality is equivalent to the pair of monotonicity statements that $\lambda_{\rm E}$ is non-increasing while $\nu^2\lambda_{\rm E}$ is non-decreasing. Comparing $\lambda_0-\lambda_{\rm E}(\nu)=\nu^2\!\int\dd\mu(\omega')/(\omega'^2+\nu^2)$ with $\lambda_2=\int\dd\mu(\omega')/\omega'^2$ over the spectral support $\omega'\ge\omega_{\rm gap}$ then brackets the dynamical Love number from both sides,
\begin{equation}
\frac{\lambda_0-\lambda_{\rm E}(\nu)}{\nu^{2}}
\ \le\ \lambda_{2}\ \le
\Big(1+\frac{\nu^{2}}{\omega_{\rm gap}^{2}}\Big)
\frac{\lambda_0-\lambda_{\rm E}(\nu)}{\nu^{2}}\,.
\label{eq:pinch}
\end{equation}
The interval contracts as $\nu\to0$ and spans at most a factor of two at
$\nu=\omega_{\rm gap}$. If the static response, the threshold, and a single value of the
Euclidean response are known, from a calculation or a dispersive reconstruction,
$\lambda_2$ is therefore confined to a narrow interval. The inequality follows directly from the
positive kernel and the support condition, and it is closely related to classical boundary
interpolation inequalities for Pick functions~\cite{Sarason:1994}.

\new{The status of the assumptions differs between regimes. In the non-relativistic description of a non-dissipative star the response is a convergent sum over normal modes with positive weights, and the positive measure, the threshold, the absence of additional subtractions and $\lambda_\infty=0$ are properties of the model, so that Eqs.~\eqref{eq:edgebound}--\eqref{eq:pinch} hold. In the fully relativistic case a global positive spectral representation of the complete matched response has not been established: the modes are damped by radiation, the response depends on the matching, and the renormalized $\lambda_2$ runs with the subtraction scale~\cite{Apostolidis:2026dtr}, which signals spectral weight extending to zero frequency. The comparison below is therefore encouraging, but it involves additional matching and subtraction assumptions and does not independently establish the bound in the relativistic regime.}

\subsection{Comparison with existing results for neutron stars}
The bounds on the dynamical neutron star  are consistent with the numerical evaluations of stellar responses found in the literature. In the Newtonian mode-sum representation the fundamental $f$-mode is typically dominant, with lower-frequency $g$-mode resonances still relevant when their tidal overlap is non-vanishing \cite{Kokkotas:1995xe,Andersson:2019}. Ideal models with $\omega_{\rm gap}=\omega_f$ and dominant
$f$-mode spectral weight nearly saturate the upper
bound~\eqref{eq:edgebound}. {\colour{blue}{Fully relativistic computations of the dynamical response for non rotating stars~\cite{Hegade:2024} also find a positive dynamical correction growing toward the $f$ mode resonance. We regard this only as qualitative agreement at the level of the full relativistic response because that calculation is not formulated as a matching to the point particle EFT and therefore does not by itself test a bound on an EFT Wilson coefficient. A quantitative comparison requires matching the relativistic response to the same worldline operators and the same subtraction prescription, which is provided by Ref.~\cite{Apostolidis:2026dtr} and is discussed below.}} The framework also explains why the single-mode approximation works so well. {\colour{blue}{
At fixed spectral contribution $\lambda_0-\lambda_\infty$, moving spectral
weight to higher frequencies necessarily decreases $\lambda_2$, as follows
from Eq.~\eqref{eq:momentbound}.
}}

A sharper, quantitative point of contact with the literature is provided by the effective spectral frequency
\begin{equation}
\omega_*\equiv\sqrt{\frac{\lambda_0}{\lambda_2}}\,,
\label{eq:omegastar}
\end{equation}
in terms of which the moment bound~\eqref{eq:edgebound} (with $\lambda_\infty=0$) is simply the statement $\omega_*\ge\omega_{\rm gap}$, with equality for the extremal single-mode response. {\colour{blue}{
The condition $\lambda_\infty=0$ in this diagnostic is appropriate to the
ideal single mode model and to a comparison performed in the same
subtraction convention.
We do not assume that a generic relativistic neutron star has a universal
vanishing value of $\lambda_\infty$.
The general stellar bound is Eq.~\eqref{eq:edgebound}, which depends on
$\lambda_0-\lambda_\infty$ and does not require
$\lambda_\infty=0$.
}} Precisely this diagnostic has been computed in the recent fully relativistic EFT matching of the neutron star dynamical tidal response~\cite{Apostolidis:2026dtr}.
{\colour{blue}{
Ref.~\cite{Apostolidis:2026dtr} performs the matching to the point particle
EFT and makes explicit that the second order dynamical Love coefficient is
a renormalized quantity whose finite part depends on the subtraction
prescription.
A finite counterterm may therefore shift the numerical value of
$\lambda_2(\mu)$, including to a negative value, without changing the
complete matched response.
The Schwarz--Pick constraint is consequently a statement about the complete
susceptibility after the subtraction data have been specified and not a
scheme independent bound on the isolated Wilson coefficient
$\lambda_2(\mu)$.
}}

\begin{figure}[t]
\centering
\includegraphics[width=\textwidth]{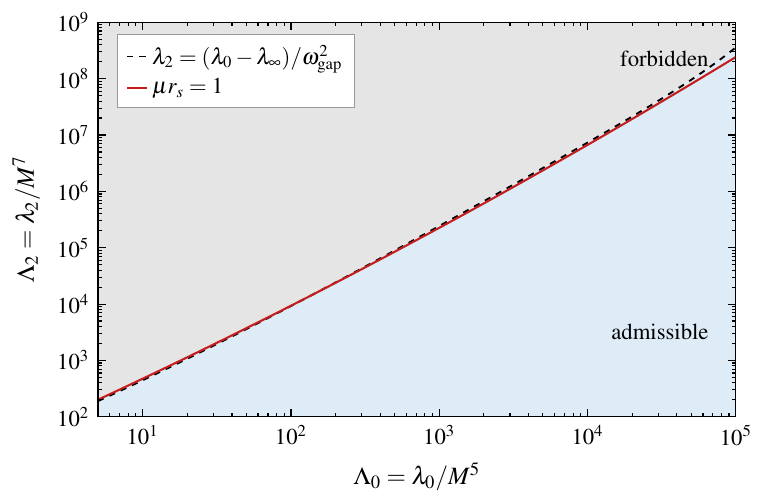}
\caption{
For $\mu r_s=1$, where $r_s=2GM$ and $M$ the neutron mass, we plot the bound for the neutron star (dashed line) versus the corresponding fit of the universal relation
derived from Eq. (61) of Ref. \cite{Apostolidis:2026dtr}.}
\label{fig:bound2}
\end{figure}

{\colour{blue}{
Within the subtraction prescriptions considered in
Ref.~\cite{Apostolidis:2026dtr}, the Newtonian value of $\omega_*$ agrees
with the $f$ mode frequency within $0.1\%$, while in the relativistic
regime the ratio $\omega_*/\omega_f$ remains close to unity over the
quoted range of renormalization scales. \color{red}{This numerical proximity is consistent with single mode dominance
in that prescription, but does not by itself establish saturation
of Eq.~\eqref{eq:edgebound}. A spectral interpretation additionally
requires that the coefficient entering $\omega_*$ be the convergent
moment of a positive spectral measure, with $\lambda_\infty=0$
and $\omega_{\rm gap}=\omega_f$.}
}}
Fig. \ref{fig:bound2} shows the corresponding between the approximate bound and the numerical fit of the universal relation for neutron stars obtained for $2G\mu M=1$, see Eq. (61) of Ref. \cite{Apostolidis:2026dtr}. The overlap is quite impressive.
{\color{red}{Finally, applying the two-sided bracket~\eqref{eq:pinch} to the
numerical relativistic responses~\cite{Hegade:2024,Apostolidis:2026dtr}
requires establishing the assumed positive-measure representation
and a strictly positive spectral threshold. Under these assumptions,
a single Euclidean value at $0<\nu\le\omega_{\rm gap}$ confines
$\lambda_2$ to an interval spanning at most a factor of two.}
}

\section{Black holes}
\label{sec:bh}
{\color{red}
{
For four-dimensional asymptotically flat Schwarzschild and Kerr black holes the
conservative static Love numbers vanish, $\lambda_0=0$
\cite{Hui:2020xxx,LeTiec:2020spy,Charalambous:2021mea,Kehagias:2024rtz,Combaluzier--Szteinsznaider:2025eoc,Gounis:2024hcm,Charalambous:2021kcz,Hui:2021vcv,Charalambous:2022rre},
and the leading response is horizon absorption, with tidal-heating coefficient
$\eta$~\cite{Chia:2020yla,Saketh:2023bul}. For the Schwarzschild quadrupole
$\eta=\tfrac{32}{45}G^5M^6$~\cite{Poisson:2004}. Unlike the ideal star, the absorptive weight
$g(\omega)\equiv\im\chi(\omega)/\omega$, in terms of which $\dd\mu=(2/\pi)g\,\dd\omega$, reaches
$\omega=0$ with $g(0)=\eta>0$. The spectrum is gapless and the inverse moments of
Eq.~\eqref{eq:moments} diverge in the infrared.

\subsection{Subtracted dispersion relation and finite-part moments}
\label{sec:subtracted}
\label{sec:finitepart}

Subtracting the Kramers--Kronig relation at $\omega=0$, with the same large semicircle of
radius $\Lambda$, the arc contributions cancel up to a remainder bounded by
$|\omega|\max_{|\omega'|=\Lambda}|\chi|/(\Lambda-|\omega|)$. Hence, provided the response grows
slower than linearly in the upper half-plane,
\begin{equation}
\max_{|\omega|=\Lambda,\ \im\omega\ge0}|\chi(\omega)|=o(\Lambda),
\label{eq:sublinear}
\end{equation}
one obtains
\begin{equation}
\re\chi(\omega)-\lambda_0=\frac{2\omega^2}{\pi}\,\mathrm P\!\!\int_0^\infty
\dd\Omega\,\frac{g(\Omega)}{\Omega^2-\omega^2},
\label{eq:subtracted}
\end{equation}
which requires neither the existence nor the value of $\chi(\infty)$. For $\lambda_0=0$ the
same hypothesis makes $G=\chi/\omega$ Herglotz and reproduces
Eqs.~\eqref{eq:Gfrommeasure}--\eqref{eq:Gpositive} without assuming a finite $\lambda_\infty$.
Since $\mathrm P\!\int_0^\infty\dd\Omega/(\Omega^2-\omega^2)=0$, the dispersive statement that
$i\eta\omega$ does not contribute to $\re\chi$, one may replace $g\to g-\eta$ in
Eq.~\eqref{eq:subtracted}. If $g-\eta=\mathcal O(\Omega^2)$ the small-frequency expansion is then
legitimate and gives
\begin{equation}
\lambda_2=\frac{2}{\pi}\int_0^\infty\dd\Omega\,\frac{g(\Omega)-g(0)}{\Omega^2},
\label{eq:lambda2fp}
\end{equation}
the finite part of the divergent moment in Eq.~\eqref{eq:moments}; the higher coefficients
follow by subtracting the Taylor polynomial of $g$. Positivity fixes $\eta$ but not the sign
of $\lambda_2$, which is set by the shape of the absorption spectrum. If
$g=\eta+\gamma|\Omega|+\dots$, Eq.~\eqref{eq:lambda2fp} diverges logarithmically and $\lambda_2$
runs, $\mu\,\dd\lambda_2/\dd\mu=-2\gamma/\pi$: the running is fixed by the first non-analytic
term of the absorption.

\subsection{Euclidean response and Schwarz--Pick}
\label{sec:bheuclid}
\label{sec:bhSP}

On the Euclidean axis Eq.~\eqref{eq:subtracted} gives, exactly for every $\nu>0$,
\begin{equation}
\Psi(\nu)=-\frac{\lambda_{\rm E}(\nu)}{\nu}=\eta+\nu\,\lambda_2(\nu),
\qquad
\lambda_2(\nu)=\frac{2}{\pi}\int_0^\infty\dd\Omega\,\frac{g(\Omega)-\eta}{\Omega^2+\nu^2}.
\label{eq:lambda2nu}
\end{equation}
This is the precise definition of the low-frequency coefficients. The Taylor coefficient
$\lambda_2=\lambda_2(0^+)$ exists when the limit exists. Otherwise only $\lambda_2(\nu)$, or a
renormalized $\lambda_2(\mu)$ in a given scheme, is meaningful, and the Euclidean frequency
acts as the infrared cutoff. When $G$ is Herglotz, Schwarz--Pick gives
\begin{equation}
-1\le\frac{\dd\ln\Psi}{\dd\ln\nu}\le1 ,
\label{eq:bhfullbound}
\end{equation}
equivalently, $-\lambda_{\rm E}$ increases and $-\lambda_{\rm E}/\nu^2$ decreases with $\nu$.
These are exact constraints on the Euclidean response at finite $\nu$, but they do not bound
$\lambda_2$. At $\nu\to0$ the logarithmic slope vanishes for any finite $\lambda_2$, because $G$
maps the boundary point $\omega=0$ to the interior value $i\eta$, while Schwarz--Pick and its
boundary (Julia--Carath\'eodory) form constrain derivatives only at points mapped to the
boundary, as happens for a star. Indeed, a deformation of $g$ in a narrow window near the
origin changes $\lambda_2$ by an arbitrary amount of either sign while leaving $\Psi$ at finite
$\nu$ essentially unchanged. Truncated relations such as $\lambda_2\gtrsim-\eta/(2\nu)$ are
therefore not bounds on $\lambda_2$, and a bound requires information on the shape of $g$.

\subsection{Schwarzschild}
\label{sec:check}

The low-frequency Schwarzschild response of Ref.~\cite{Combaluzier--Szteinsznaider:2025eoc}
gives
\begin{equation}
\lambda_{2,\ell}(\mu)=\eta_\ell\,r_s\left[c_\ell-\ln(\mu r_s)\right],
\label{eq:CGJRSresult}
\end{equation}
with scheme-dependent constants $c_\ell$. As an illustrative
RG-improved prescription compatible with this running and retarded
analyticity, we consider,
\begin{equation}
\chi_\ell(\omega)\supset\eta_\ell\,r_s\,\omega^2\left[c_\ell-\ln(-i\omega r_s)\right],
\qquad
\lambda_{2,\ell}(\nu)=\eta_\ell\,r_s\left[c_\ell-\ln(\nu r_s)\right]+\mathcal O(\nu).
\label{eq:Schlog}
\end{equation}
This frequency dependence is an additional assumption and does not
follow from the running alone. Under this prescription,
$\gamma=\tfrac{\pi}{2}\eta_\ell r_s$, corresponding to an absorptive
weight $g(\omega)$ rising linearly in $|\omega|$ away from $\eta_\ell$.
An independent derivation of this term from the absorption spectrum
of the same matched response has not been carried out.

\subsection{Status of the black hole results}
\label{sec:bhscope}

The statements $\eta\ge0$ and the absence of any positivity bound on $\lambda_2$ are exact.
Eqs.~\eqref{eq:subtracted}--\eqref{eq:bhfullbound} are conditional on Eq.~\eqref{eq:sublinear}
for the complete matched response, and Eq.~\eqref{eq:Schlog} is illustrative. The growth
hypothesis is not established. The closed-form scheme-independent Schwarzschild
response~\cite{Solon:2026clf}, continued literally to complex frequency, grows faster than any
power in the sector $\pi/4<\arg\omega<3\pi/4$ because of its far-zone Coulomb dressing, and
therefore cannot be the complete response there. Whether the complete matched response
satisfies Eq.~\eqref{eq:sublinear} is open. The physical conclusion is that gapless horizon
absorption is precisely what prevents positivity from bounding the dynamical Love number,
while a positive spectral threshold is what converts positivity into a bound.
}}

\section{Conclusions}
\label{sec:conclusions}
We have derived Schwarz--Pick constraints on the frequency-dependent tidal
response of compact objects.  Whenever the complete matched susceptibility,
including its subtraction data, can be rescaled into an axis-preserving
Herglotz function, the Schwarz--Pick theorem gives the exact logarithmic-slope
inequality~\eqref{eq:master}.  This is a constraint on the complete Euclidean
response and does not, without additional information, provide separate bounds
on scheme-dependent coefficients in a low-frequency expansion.

For a stellar response admitting the positive-measure representation of
Section~\ref{sec:response}, \new{with convergent inverse moments,} positivity fixes the sign
$\lambda_2\geq0$.  If the tidal spectral measure has a strictly positive lower
endpoint $\omega_{\rm gap}$, it also gives the sharp moment bound
\eqref{eq:edgebound} and the two-sided finite-frequency interval
\eqref{eq:pinch}.  The ideal single-mode response saturates the moment bound.
The relevant scale is the lowest frequency carrying non-zero tidal spectral
weight, which need not coincide with the most prominent mode. \new{These are theorems for the
ideal non-relativistic star, while the relativistic comparison involves additional matching
and subtraction assumptions. For gapless dissipative stars the finite bound is unavailable and
the sign of $\lambda_2$ is not fixed.}

\new{For a black hole, passivity gives $\eta\geq0$ and the spectrum is gapless. If the
complete response grows slower than linearly at large complex frequency, a dispersion relation
subtracted at zero frequency makes $\chi/\omega$ Herglotz, defines $\lambda_2$ as a finite-part
moment of the absorption spectrum, relates its running to non-analytic absorption, and
Schwarz--Pick constrains the Euclidean response, Eq.~\eqref{eq:bhfullbound}. It does not bound
$\lambda_2$. The growth hypothesis is not established, and the literal continuation of the
scheme-independent Schwarzschild skeleton violates it.}

Two assumptions delimit the scope of the results.  First, non-negative
absorption on the real axis does not by itself imply the interior Herglotz
property, since the required high-frequency behavior and subtraction data must also
be controlled.  Second, the finite stellar moment bound requires a strictly
positive spectral threshold.  Even when this threshold is absent, the
full-function Schwarz--Pick inequality remains applicable whenever the
corresponding complete Herglotz representative can be established.

\section*{Acknowledgements}
We thank Valerio  De Luca and Davide Perrone for useful comments and discussions. A.R.\ acknowledges support from the Swiss National Science Foundation (project number
CRSII5\_213497).

    \appendix

\section{Herglotz functions and the Schwarz-Pick theorem}
\label{secHerglotz}

This appendix summarizes the geometric properties of Herglotz functions and derives the logarithmic slope bound used in the main text. 

A Herglotz function $F$ is a holomorphic map from the open upper half-plane $\mathbb{H}$ to the closed upper half-plane, meaning that $\operatorname{Im} F(w) \ge 0$ for all $w \in \mathbb{H}$. For any non-constant physical response, the minimum principle for harmonic functions forces the strict inequality $\operatorname{Im} F(w) > 0$ throughout the interior. 
Through the Herglotz representation theorem, such a function can be written as
\begin{equation}
F(w) = \alpha + \beta w + \int_{\mathbb{R}} \left( \frac{1}{t-w} - \frac{t}{1+t^2} \right) \mathrm{d}\rho(t),
\label{eqherglotz}
\end{equation}
where $\alpha$ is real, $\beta \ge 0$, and $\rho$ is a positive Borel measure. This is the mathematical statement of the Kramers-Kronig dispersion relation for a passive, causal system. In particular, the positivity of the spectral measure $\rho$ is the direct manifestation of physical passivity.

To find the constraints on the imaginary axis, we use the Schwarz-Pick theorem. This theorem states that any holomorphic self-map of the upper half-plane $\mathbb{H}$ is a contraction with respect to its natural hyperbolic metric
\begin{equation}
\mathrm{d}s^2 = \frac{|\mathrm{d}w|^2}{(\operatorname{Im} w)^2}.
\label{eqhpmetric}
\end{equation}
The hyperbolic distance between two points $w_1$ and $w_2$ is given by
\begin{equation}
d_{\mathbb{H}}(w_1, w_2) = \operatorname{arccosh}\left( 1 + \frac{|w_1 - w_2|^2}{2 \operatorname{Im} w_1 \operatorname{Im} w_2} \right).
\label{eqhpdist}
\end{equation}
Under any Herglotz function $F$, the Schwarz-Pick theorem states that this distance obeys the contraction property
\begin{equation}
d_{\mathbb{H}}(F(w_1), F(w_2)) \le d_{\mathbb{H}}(w_1, w_2).
\label{eqspfinite}
\end{equation}

We now evaluate this constraint along the imaginary frequency axis, where $w = i\nu$ with $\nu > 0$. For a real physical response, $F(i\nu)$ must be purely imaginary, so we write $F(i\nu) = i \Phi(\nu)$ with $\Phi > 0$. 
Along this axis, the hyperbolic metric simplifies to $\mathrm{d}s = \mathrm{d}\nu/\nu$, meaning the distance is simply the difference of the logarithms
\begin{equation}
d_{\mathbb{H}}(i\nu_1, i\nu_2) = \left| \ln\nu_1 - \ln\nu_2 \right|.
\label{eqlogdist}
\end{equation}
Applying the contraction property to $F(i\nu)$ yields
\begin{equation}
\left| \ln\Phi(\nu_1) - \ln\Phi(\nu_2) \right| \le \left| \ln\nu_1 - \ln\nu_2 \right|
\label{eqtwopoint}
\end{equation}
When the two points merge, the global contraction yields the local master inequality
\begin{equation}
\left| \frac{\mathrm{d}\ln\Phi}{\mathrm{d}\ln\nu} \right| \le 1 ,
\label{eqmaster}
\end{equation}
the statement that the logarithmic slope of the Euclidean response cannot exceed unity in magnitude, a direct consequence of the hyperbolic geometry imposed by causality and passivity.

\section{The ideal mode sum, the low-frequency expansion, and the spectral threshold for neutron stars}
\label{app:modesum}
The dynamical Love numbers are usually defined by a low-frequency
Taylor expansion. Here we explain  how such an  expansion is related to the stellar mode spectrum, and why the
lower endpoint of the tidally coupled spectral measure appears in the finite moment bound.
The identification of this threshold with the radius of convergence is exact in the ideal
undamped meromorphic model described here \cite{Pitre:2023xsr}. In a dissipative system with complex poles,
branch cuts, or continuous low-frequency spectral weight, the spectral threshold and the
distance to the nearest complex singularity are distinct notions.

Linear tidal perturbations of an ideal non-rotating star decompose into normal modes
labelled by $n$, with real frequencies $\omega_n$ and mass-multipole overlaps
$\langle Q_n\rangle$. In a fixed multipolar channel, summing their driven responses gives
\begin{equation}
\lambda(\omega)=\lambda_\infty+
\sum_n\frac{\lambda_n\,\omega_n^{2}}{\omega_n^{2}-\omega^{2}},
\qquad
\lambda_n=\frac{|\langle Q_n\rangle|^{2}}{\omega_n^{2}}\ge0,
\label{eq:app-modesum}
\end{equation}
where each undamped mode contributes poles at $\omega=\pm\omega_n$. 
The representation above should be understood as an idealized
positive-weight mode model. 
In the folded
positive-frequency representation used in the main text, the corresponding measure is
\begin{equation}
\dd\mu(\omega')=\sum_n\lambda_n\,
\delta(\omega'-\omega_n)\,\dd\omega'\ge0,
\label{eq:app-discretemeasure}
\end{equation}
and the positivity of the measure is the positivity of the squared tidal overlaps.

Notice that in full general relativity, the construction
of a mode-sum response is more subtle because the relativistic mode
inner product need not be positive definite on the complete perturbation
space, and formal convergence of the full mode expansion is not
automatic. Nevertheless, the dominant $f$-mode contribution has been
found to reproduce the directly matched relativistic response with good
accuracy for non-rotating stellar models
\cite{HegadeKR:2026kku}.

For frequencies smaller than the lowest singularity of the response,
each modal contribution can be expanded as a convergent geometric series,
\begin{equation}
    \frac{\omega_n^2}{\omega_n^2-\omega^2}
    =
    \sum_{k=0}^{\infty}
    \left(\frac{\omega}{\omega_n}\right)^{2k},
    \qquad
    |\omega|<\omega_n .
\end{equation}
Substituting this expansion into the mode sum gives
\begin{equation}
    \lambda(\omega)
    =
    \lambda_0
    +
    \lambda_2\omega^2
    +
    \lambda_4\omega^4
    +\cdots ,
\end{equation}
where
\begin{equation}
    \lambda_0
    =
    \lambda_\infty+\sum_n\lambda_n,
    \qquad
    \lambda_{2k}
    =
    \sum_n\frac{\lambda_n}{\omega_n^{2k}}
    =
    \int_0^\infty
    \frac{\dd\mu(\omega')}{\omega'^{2k}},
    \qquad
    k\geq1 .
    \label{eq:app-moments}
\end{equation}
The low-frequency expansion therefore does not erase the information about the stellar
modes. The mode frequencies and tidal overlaps are encoded in the coefficients
$\lambda_{2k}$. Each mode contributes to
every coefficient, but with a weight proportional to
$\omega_n^{-2k}$. Consequently, higher-order coefficients are increasingly
sensitive to low-frequency modes. A low-frequency $g$-mode~\cite{Lai:1993di} or inertial
mode may therefore make only a small contribution to the static response
$\lambda_0$, while contributing much more significantly to the dynamical
coefficient $\lambda_2$ because of the additional factor
$1/\omega_n^2$.

In the ideal model~\eqref{eq:app-modesum}, define the lowest tidally coupled frequency by
\begin{equation}
    \omega_{\rm gap}
    \equiv
    \min\left\{
        \omega_n \,:\, \lambda_n\neq 0
    \right\}.
    \label{eq:app-gap}
\end{equation}
In other words, $\omega_{\rm gap}$ is the frequency of the lowest normal mode with
non-vanishing tidal overlap. Modes with $\lambda_n=0$ do not contribute
to the tidal response and are therefore excluded from the minimum.
Because the only singularities are the real poles $\pm\omega_n$, the Taylor series has
radius of convergence $\omega_{\rm gap}$. In a simple cold barotropic star this frequency
is commonly the quadrupolar $f$-mode and may be denoted $\omega_f$. In a more general star
it is the lowest mode with non-zero overlap, irrespective of whether that mode is dominant.
With damping, a continuum, or non-meromorphic contributions, the convergence radius is set
by the nearest complex singularity and need not equal the lower endpoint of the positive
dissipation measure.

On the Euclidean axis, obtained by setting $\omega=i\nu$ with $\nu>0$,
the ideal mode-sum response becomes
\begin{equation}
    \lambda_{\rm E}(\nu)
    \equiv
    \lambda(i\nu)
    =
    \lambda_\infty
    +
    \sum_n
    \frac{\lambda_n\omega_n^2}
         {\omega_n^2+\nu^2}
    =
    \lambda_\infty
    +
    \int_0^\infty
    \frac{\omega'^2}
         {\omega'^2+\nu^2}\,
    \dd\mu(\omega') .
    \label{eq:app-euclid}
\end{equation}
Each term in the sum is real and decreases as $\nu$ increases, so
$\lambda_{\rm E}(\nu)$ is real and monotonically decreasing, interpolating between
\begin{equation}
    \lambda_{\rm E}(0)=\lambda_0
\end{equation}
and
\begin{equation}
    \lambda_{\rm E}(\nu)\longrightarrow\lambda_\infty
    \qquad
    \text{as }
    \nu\longrightarrow\infty .
\end{equation}
The upper bound on the dynamical coefficient follows directly from the
fact that every mode contributing to the response satisfies
$
    \omega_n\geq\omega_{\rm gap}$,
and hence
\begin{align}
    \lambda_2
    =
    \sum_n\frac{\lambda_n}{\omega_n^2}
    \leq
    \frac{1}{\omega_{\rm gap}^2}
    \sum_n\lambda_n=
    \frac{\lambda_0-\lambda_\infty}
         {\omega_{\rm gap}^2},
    \label{eq:app-cs}
\end{align}
which is precisely the bound in Eq.~\eqref{eq:edgebound}.

The derivation uses only the positivity of the modal weights and the existence of a
lowest tidally coupled frequency. It never evaluates the truncated low-frequency
Schwarz--Pick expansion at $\nu=\omega_{\rm gap}$.

The inequality is saturated when all the spectral weight is concentrated
in a single mode at the threshold frequency,
$\omega_n=\omega_{\rm gap}$. Moving part of the spectral weight to higher-frequency modes at fixed
$\lambda_0-\lambda_\infty=\sum_n\lambda_n$ shrinks the factors $1/\omega_n^2$ and hence
$\lambda_2$. This is why an ideal stellar response dominated by a single $f$-mode lies
close to the upper bound.

\bibliographystyle{JHEP}
\bibliography{refs}

\end{document}